\documentclass[twocolumn,preprintnumbers,amsmath,amssymb]{revtex4}
\usepackage{amssymb}
\usepackage{latexsym}
\usepackage{epsfig}
\usepackage{color}

\begin{document}

\title{Generalized phenomenological models of dark energy}

\author{Prasenjit Paul}
\email{prasenjit071083@gmail.com}
\affiliation{Department of Physics, Indian Institute of Engineering Science and Technology, Shibpur, India}

\author{Rikpratik Sengupta}
\email{rikpratik.sengupta@gmail.com}
\affiliation{Department of Physics, Government College of Engineering and Ceramic Technology, Kolkata 700010, West Bengal, India}

\date{\today}

\begin{abstract}
It was first observed at the end of the last century that the universe is presently accelerating. Ever since, there have been several attempts to explain this observation theoretically. There are two possible approaches. The more conventional one is to modify the matter part of the Einstein Field Equations and the second one is to modify the geometry part. We shall consider two phenomenological models based on the former, more conventional approach within the context of General Relativity. The phenomenological models in this paper consider a $\Lambda$ term as a function of $\frac{\ddot{a}}{a}$ and $\rho$ where $a$ and $\rho$ are the scale factor and matter-energy density respectively. Constraining the free parameters of the models with latest observational data gives satisfactory values of parameters as considered by us initially. Without any field-theoretic interpretation, we explain the recent observations with a dynamical cosmological constant.    
\end{abstract}


\maketitle

\section{Introduction}

Type Ia high redshift supernova observations indicate that the universe is presently accelerating \cite{perlmutter,riess}. This is mostly thought to be due to the presence of some unknown fluid known as Dark Energy. Soon after the first cosmological solution to the Einstein Field Equations (EFE), Einstein had put an additional $\Lambda$ term (known as the cosmological constant) which produced a repulsive effect,in order to modify the EFE so that the cosmological solution could lead to a static universe. He later called the introduction of this term to be the greatest blunder of his life. However after observations suggested an accelerating universe, there was a revived interest in the $\Lambda $ term as a possible candidate for the Dark energy. Theoretically, cosmological constant is assumed to be contribution from vacuum energy given by $\Lambda = 8 \pi G \rho_{vac},$ arising out of quantum vacuum fluctuations of some fundamental field. Although the calculated value of $\rho_{vac}$ turns out to be much larger than the value  of $\Lambda$ predicted from observations, but there is no theoretical argument of making the $\rho_{vac}$ term vanish to exactly zero \cite{aguirre}. So $\Lambda$ models are favored for Dark energy (DE). $\Lambda$ has also been thought to be generated from particle creation effect or dynamical scalar field \cite{sahni}. If we consider that $\Lambda$ term is responsible for the dark energy, whatever be the generation mechanism, it is clear that contrary to Einstein, $\Lambda$ is not a constant but a dynamical cosmological term \cite{overduin}. 

DE is also some times considered without the presence of any fluid or $\Lambda$ term, just as a consequence of the modification of the geometric part or the left hand side of the EFE, but such efforts are not possible in the context of standard General Relativity (GR) \cite{Lue,Nojiri}. There are also dynamically evolving scalar field models which have been used to describe DE. The popular dynamical physical field models that have been utilized for this purpose are quintessence \cite{zlatev,brax,barreiro,albrecht}, K-essence \cite{mukhanov1,mukhanov2,chiba,steinhardt1,steinhardt2,liddle}, phantom \cite{cladwell} and tachyonic field \cite{sen1,sen2,garousi1,garousi2,garousi3,bergshoeff,kluson,kutasov}. Phenomenological models of a dynamical $\Lambda$ term are also being popularly considered as candidates of DE. Phenomenological simply means there is no derivation of the dynamical $\Lambda$ term from any underlying quantum field theory. Such models may be categorized into three types: (i) kinematic (ii) hydrodynamic and (iii) field theoretic. The first means $\Lambda$ is a function of time or scale factor $a(t)$. The second means $\Lambda$ is treated as a barotropic fluid with some Equation of State (EOS). The third means $\Lambda$ is treated as a new physical classical field with a phenomenological Lagrangian. We will be concerned here with (i) and (ii) only. Such kinematic and hydrodynamic models have been treated in some depth before. A dynamical model with $\Lambda=\alpha H^2$, where $H(t)=\frac{\dot{a}}{a}$ has been explored by Mukhopadhyay et al. \cite{Mukhopadhyay}. A similar model with $\dot{\Lambda} \sim H^3$ has been considered in \cite{ray3,usmani}. 

The most frequently used forms of $\Lambda$ for phenomenological models are $\Lambda=\alpha \big(\frac{\dot{a}}{a}\big)^2$, $\Lambda= \beta \frac{\ddot{a}}{a}$ and $\Lambda=8 \pi G \gamma \rho,$ where $\alpha$, $\beta$ and $\gamma$ are constants whose values can be constrained from observations. The first type of model has been considered by \cite{carvalho,waga,lima,salim,arbab1,wetterich,arbab,padmanabhan}. The second model has been dealt with by \cite{arbab2,arbab3,arbab4,overduin}. The third type of model has been considered by \cite{vishwakarma}. The equivalence of these three forms has been shown by Ray et. al. \cite{ray1,ray2}, connecting the free parameters of the models with the matter density and vacuum energy density parameters in the first paper and by application of numerical methods in the later one. This paper is basically an in-depth extension of the work done by Mukhopadhyay et. al. \cite{mukhopadhyay2} where they have considered the first type of model and obtained cosmological solutions for any possible value of the curvature constant and equation of state papameter $\omega$. They have also analysed the physical features of the solutions. We shall do the same for the second and third models and also compare our results to the latest observational data constraining our free parameters. The constraints are found to be exactly compatible with our initial considerations.

The paper is organized as follows. In the second section we consider the mathematical model in the background of an isotropic FLRW space-time based on GR. We calculate the various cosmological and physical parameters for the two different phenomenological models in consideration. In the next section we constrain the free parameters associated with the models based on recent observational data. The final section summarizes the physical insights of the results we have obtained.

\section{Mathematical model}
The Einstein field equation (EFE) in presence of a cosmological constant, $\Lambda(t)$ is given by
\begin{equation}
  G^{\mu\nu}=-8\pi G\Bigg[T^{\mu\nu}-\frac{\Lambda}{8\pi G}g^{\mu\nu}\bigg],
\end{equation}
where we shall take the cosmological constant as a function of time in order to account for the dark energy. We obtain the EFE for the cosmological FLRW metric
\begin{equation}
ds^2=-dt^2+\bigg[\frac{dr^2}{1-kr^2}+r^2(d\theta^2+\sin^2\theta d\phi^2)\bigg],
\end{equation}
 which yields the equations 
 \begin{equation}
 \bigg(\frac{\dot{a}}{a}\bigg)^2+\frac{k}{a^2}=\frac{8\pi G\rho}{3}+\frac{\Lambda}{3},
 \end{equation}

\begin{equation}
 \bigg(\frac{\ddot{a}}{a}\bigg)=-\frac{4\pi G(\rho+3p)}{3}+\frac{\Lambda}{3},
 \end{equation}
where $a(t)$ and $k$ are scale factor and curvature constant respectively.

The energy-momentum conservation gives
\begin{equation}
8\pi G(p+\rho)\bigg(\frac{\dot{a}}{a}\bigg)=-\frac{8\pi G}{3}\dot{\rho}-\frac{\dot{\Lambda}}{3}.
\end{equation}

We consider barotropic fluid with Equation of state (EOS) of the form 
\begin{equation}
p=\omega \rho,
\end{equation}
where $\omega$ denotes the EOS parameter which can assume specific values during the evolution of the Universe for different phases.  Plugging this relation in Eq. (4), the energy density is obtained as
\begin{equation}
\rho=\frac{3}{4\pi G(1+3 \omega)}\bigg(\frac{\Lambda}{3}-\frac{\ddot{a}}{a}\bigg).
\end{equation}

Substituting Eq. (6) in Eq. (4) multiplied by a factor of $\frac{2}{1+3\omega}$ and adding Eq. (3) to it we get the differential equation
\begin{equation}
\bigg(\frac{\dot{a}}{a}\bigg)^2+\frac{k}{a^2}+ \frac{2}{1+3\omega}\bigg(\frac{\ddot{a}}{a}\bigg)=\bigg(\frac{1+\omega}{1+3\omega}\bigg)\Lambda.
\end{equation}

\subsection{Solutions for Phenomenological Model $\Lambda\sim\frac{\ddot{a}}{a}$}
In this phenomenological model we consider $\Lambda=\beta\frac{\ddot{a}}{a}$, where $\beta<0$ which is justified in the light of latest observational data \cite{planck} as shown in Section (III). Using this form of $\Lambda$ in Eq. (8) we obtain
\begin{equation}
\frac{\ddot{a}}{\dot{a}}=-\frac{(1+3\omega)}{2-(1+\omega)\beta}\frac{\dot{a}}{a}-\frac{(1+3\omega)k}{2-(1+\omega)\beta}\frac{1}{a\dot{a}}.
\end{equation}  

This equation can be simplied to
\begin{equation}
a\dot{a}\frac{d}{dt}\Bigg[ln(\dot{a}a^{-A})\bigg]= Ak,
\end{equation}
where $A=-\frac{1+3\omega}{2-(1+\omega)\beta}$. We choose $A=-1$, such that $\omega=\frac{1-\beta}{3+\beta}$.

The above equation now takes the form
\begin{equation}
a\dot{a}\frac{d}{dt}\Bigg[ln(\dot{a}a)\bigg]= -k.
\end{equation}

The scale factor turns out to be
\begin{equation}
a(t)=\sqrt{A_0t+A_1-kt^2},
\end{equation} 
where $A_0$ and $A_1$ are integration constants.

As we are considering a universe evolving from a singularity, $a(t=0)=0.$ This gives $A_1=0.$ So
\begin{equation}
a(t)=\sqrt{A_0t-kt^2}.
\end{equation}

The Hubble parameter is computed as 
\begin{equation}
H(t)=\frac{A_0-2kt}{2(A_0t-kt^2)}.
\end{equation}

The cosmological constant is given by
\begin{equation}
\Lambda(t)=-\frac{\beta {A_0}^2}{4(A_0t-kt^2)^2}.
\end{equation}

The energy density is given by
\begin{equation}
\rho(t)=\frac{(3-\beta)}{16 \pi G}\frac{A_0^2}{(A_0t-kt^2)^2}.
\end{equation}

\begin{figure}
	    \includegraphics[width=1.20\linewidth]{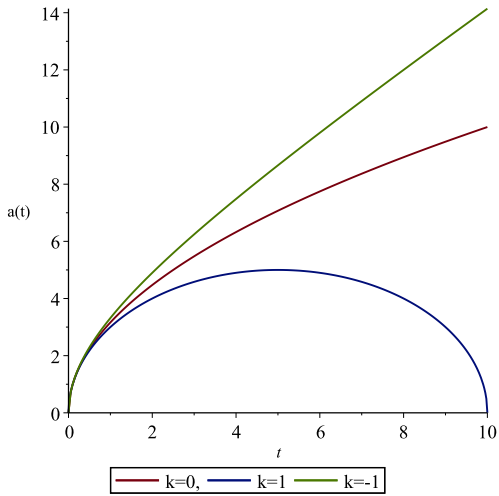}
	    \vspace{-6cm}
	     \caption{plot of $a(t)$ versus $t$ for different values of $k$.}
	\end{figure}

\begin{figure}
	\begin{center}
		\includegraphics[width=1.20\linewidth]{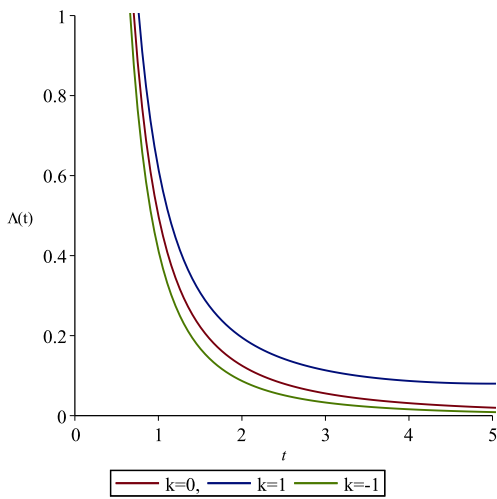}
		\vspace{-6cm}
		\caption{plot of $\Lambda(t)$ versus $t$ for different values of $k$.}
	\end{center}
\end{figure}

The variation of the scale factor and cosmological constant with time has been plotted for $k=0$, $\pm{1}$ in Figs. 1 and 2.

We obtain a closed universe for $k=1$ and open universe for $k=-1$ as expected.  

The density parameters for matter, cosmological constant and curvature respectively, can be computed for this phenomenological model as
\begin{equation}
\Omega_{m}= \frac{8 \pi G \rho}{3 H^2}=-\frac{4 \beta}{3}\Bigg[\frac{k(A_0t-kt^2)}{(A_0-2kt^2)^2}+\frac{1}{4}\Bigg],
\end{equation}

\begin{equation}
\Omega_{\Lambda}= \frac{\Lambda}{3 H^2}=\frac{4 (3+\beta)}{3}\Bigg[\frac{k(A_0t-kt^2)}{(A_0-2kt^2)^2}+\frac{1}{4}\Bigg],
\end{equation}

\begin{equation}
\Omega_{k}= -\frac{k}{a^2 H^2}=-\frac{4k(A_0t-kt^2)}{(A_0-2kt^2)^2}.
\end{equation}

For flat space ($k=0$) we see from the above expressions that the sum total of the density parameters of the above components is equal to 1, such that $\Omega_{k}=0$, $\Omega_{\Lambda}=-\frac{\beta}{3}$ and $\Omega_{m}=\frac{(3+\beta)}{3}$.

Also for both the limiting cases, $t \rightarrow 0$ and $\infty$, the sum total  of the density parameters are equal to 1. In these two cases both $\Omega_{m}$ and $\Omega_{\Lambda}$ become independent of $k$. Hence both for early and late times Universe exhibit similar behaviour as per as the $k$ dependency of $\Omega_{m}$ and $\Omega_{\Lambda}$ is concerned. 

In general, it can be observed on taking the sum of Eqs. (17)-(19) that
\begin{equation}
\Omega_{m}+\Omega_{\Lambda}+\Omega_{k}=1.
\end{equation}

The variation of the density parameters for both open and closed universes are given in Figs. 3 and 4 respectively. This analytical approach is consistent with the observational constraints $\Omega=1 \pm 0.016$ \cite{planck}.

\begin{figure}
	\begin{center}
		\includegraphics[width=1.20\linewidth]{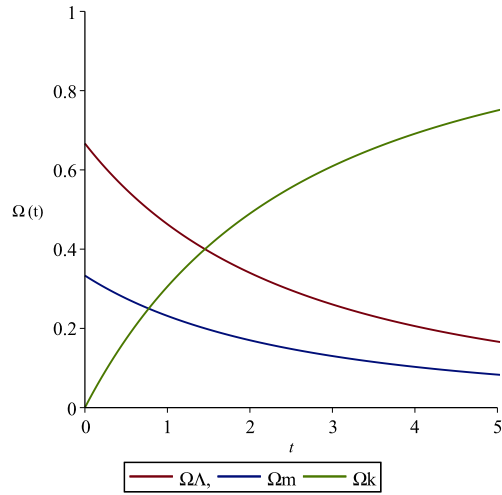}
		\vspace{-6cm}
		\caption{plot of $\Omega(t)$ versus $t$ for $k=-1$.}
	\end{center}
\end{figure}

\begin{figure}
	\begin{center}
		\includegraphics[width=1.20\linewidth]{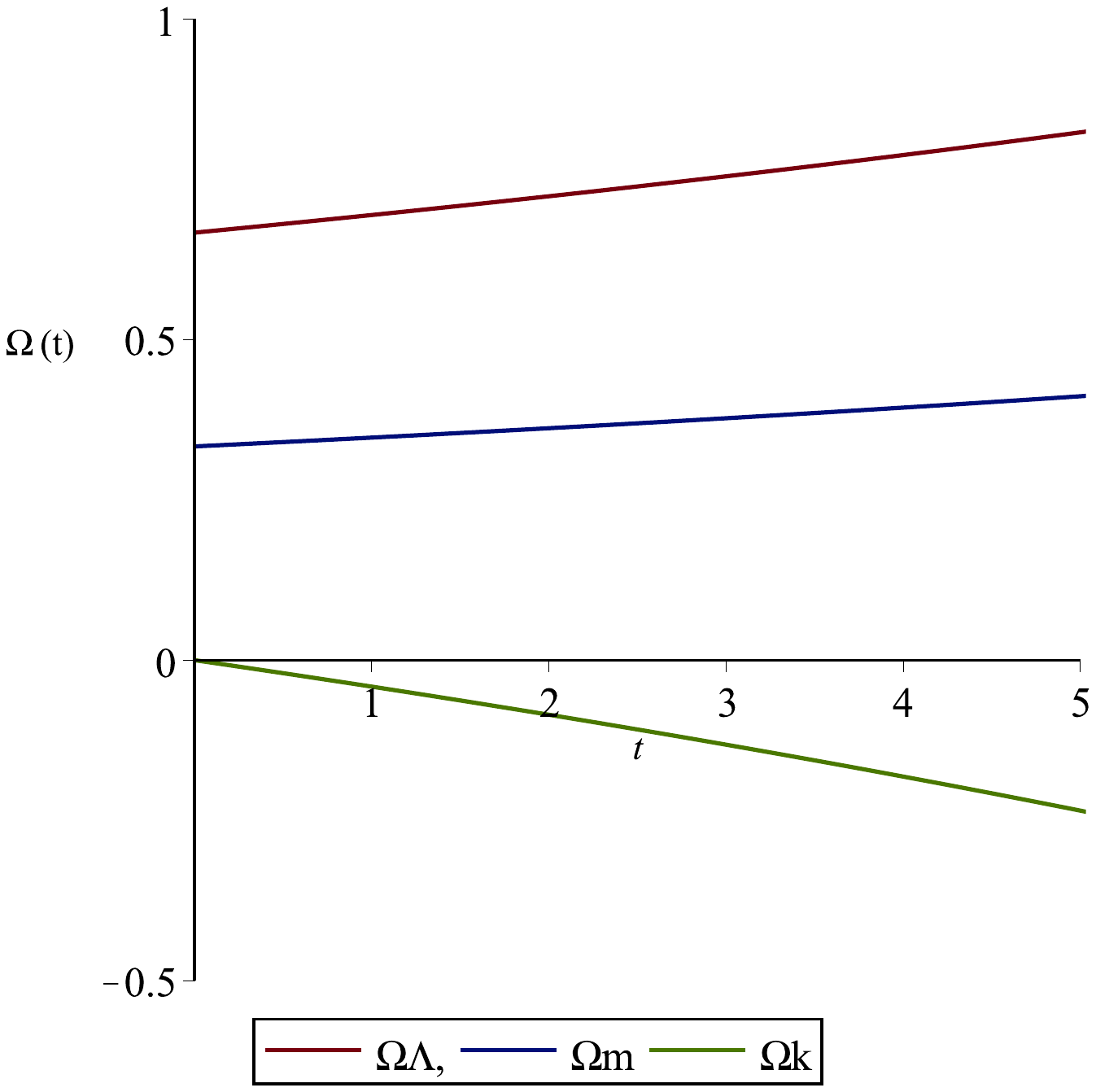}
		\vspace{-6cm}
		\caption{plot of $\Omega(t)$ versus $t$ for $k=+1$.}
	\end{center}
\end{figure}

\subsection{Solutions for Phenomenological Model $\Lambda\sim\rho$}
In this phenomenological model we consider $\Lambda=8\pi G\rho\gamma$, where $\gamma>0$ which is consistent with the observation as can be seen in Section (III). Using this form of $\Lambda$ in Eq. (8) we obtain
\begin{equation}
\frac{\ddot{a}}{\dot{a}}=-\frac{(1+3\omega-2\gamma)}{2(1+\gamma)}\frac{\dot{a}}{a}-\frac{(1+3\omega-2\gamma)}{2(1+\gamma)}\frac{k}{a\dot{a}}
\end{equation}  

This equation can be simplified to
\begin{equation}
a\dot{a}\frac{d}{dt}\Bigg[ln(\dot{a}a^{-B})\bigg]= Bk,
\end{equation}
where $B=-\frac{(1+3\omega-2\gamma)}{2(1+\gamma)}$. We choose $B=-1$, such that $\omega=\frac{4\gamma+1}{3}$.

The above equation now takes the form
\begin{equation}
a\dot{a}\frac{d}{dt}\Bigg[ln(\dot{a}a)\bigg]= -k.
\end{equation}

The scale factor turns out to be
\begin{equation}
a(t)=\sqrt{{A_0}^\prime t+A_1^\prime-kt^2},
\end{equation} 
where ${A_0}^\prime$ and ${A_1}^\prime$ are integration constants.

As we are considering a universe evolving from a singularity, $a(t=0)=0.$
This gives $A_1^{\prime}=0.$ So
\begin{equation}
a(t)=\sqrt{A_0^{\prime}t-kt^2}.
\end{equation}

The Hubble parameter is computed as 
\begin{equation}
H(t)=\frac{A_0^\prime-2kt}{2(A_0^{\prime}t-kt^2)}.
\end{equation}

The cosmological constant is given by
\begin{equation}
\Lambda(t)=-\frac{3\gamma {A_0^\prime}^2}{2(1+3\omega-2\gamma)(A_0^\prime t-kt^2)^2}.
\end{equation}

The energy density is given by
\begin{equation}
\rho(t)=\frac{3}{16\pi G(1+3\omega-2\gamma)}\frac{{A_0^\prime}^2}{(A_0^\prime t-kt^2)^2}.
\end{equation}

Variation of the scale factor $a(t)$ is same as shown in Fig. 1.

\begin{figure}
	\begin{center}
		\includegraphics[width=1.20\linewidth]{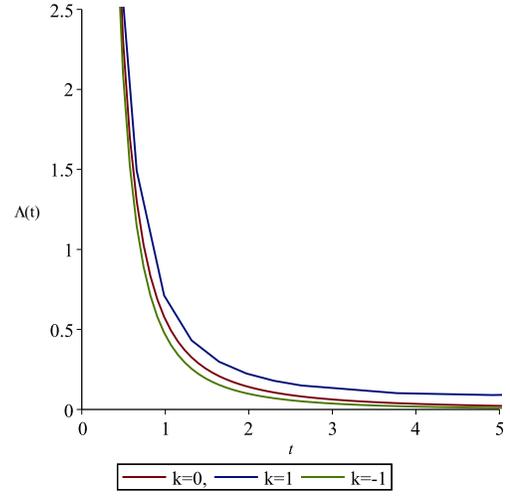}
		\vspace{-6cm}
		\caption{plot of $\Lambda(t)$ versus $t$ for different values of $k$.}
	\end{center}
\end{figure} 

The variation of the cosmological constant with time has been plotted for $k=0$, $\pm{1}$ in Fig. 5.

The density parameters for matter, cosmological constant and curvature respectively, can be computed in a similar manner as above for this phenomenological model as
\begin{equation}
\Omega_{m}= \frac{4\gamma}{1+\gamma}\Bigg[\frac{k(A_0^\prime t-kt^2)}{(A_0^\prime-2kt^2)^2}+\frac{1}{4}\Bigg],
\end{equation}

\begin{equation}
\Omega_{\Lambda}= \frac{4}{1+\gamma}\Bigg[\frac{k(A_0^\prime t-kt^2)}{(A_0^\prime-2kt^2)^2}+\frac{1}{4}\Bigg],
\end{equation}

\begin{equation}
\Omega_{k}= -\frac{k}{a^2 H^2}=-\frac{4k(A_0^\prime t-kt^2)}{(A_0^\prime-2kt^2)^2}.
\end{equation}

For flat space ($k=0$) we see from the above expressions that the sum total of the density parameters of the above components is equal to 1, such that $\Omega_{k}=0$, $\Omega_{m}=\frac{1}{1+\gamma}$ and $\Omega_{\Lambda}=\frac{\gamma}{1+\gamma}$.

In case of $t \rightarrow 0$ and $\infty$, the sum total  of the density parameters are equal to 1 in a similar manner as for the previous model. 

In general, it can be observed on taking the sum of the above three equations that 
\begin{equation}
\Omega_{m}+\Omega_{\Lambda}+\Omega_{k}=1,
\end{equation}
which is again consistent with \cite{planck}.
 
The variation of the density parameters for both open and closed Universes are given in Figs. 6 and 7 respectively. 

\begin{figure}
	\begin{center}
		\includegraphics[width=1.20\linewidth]{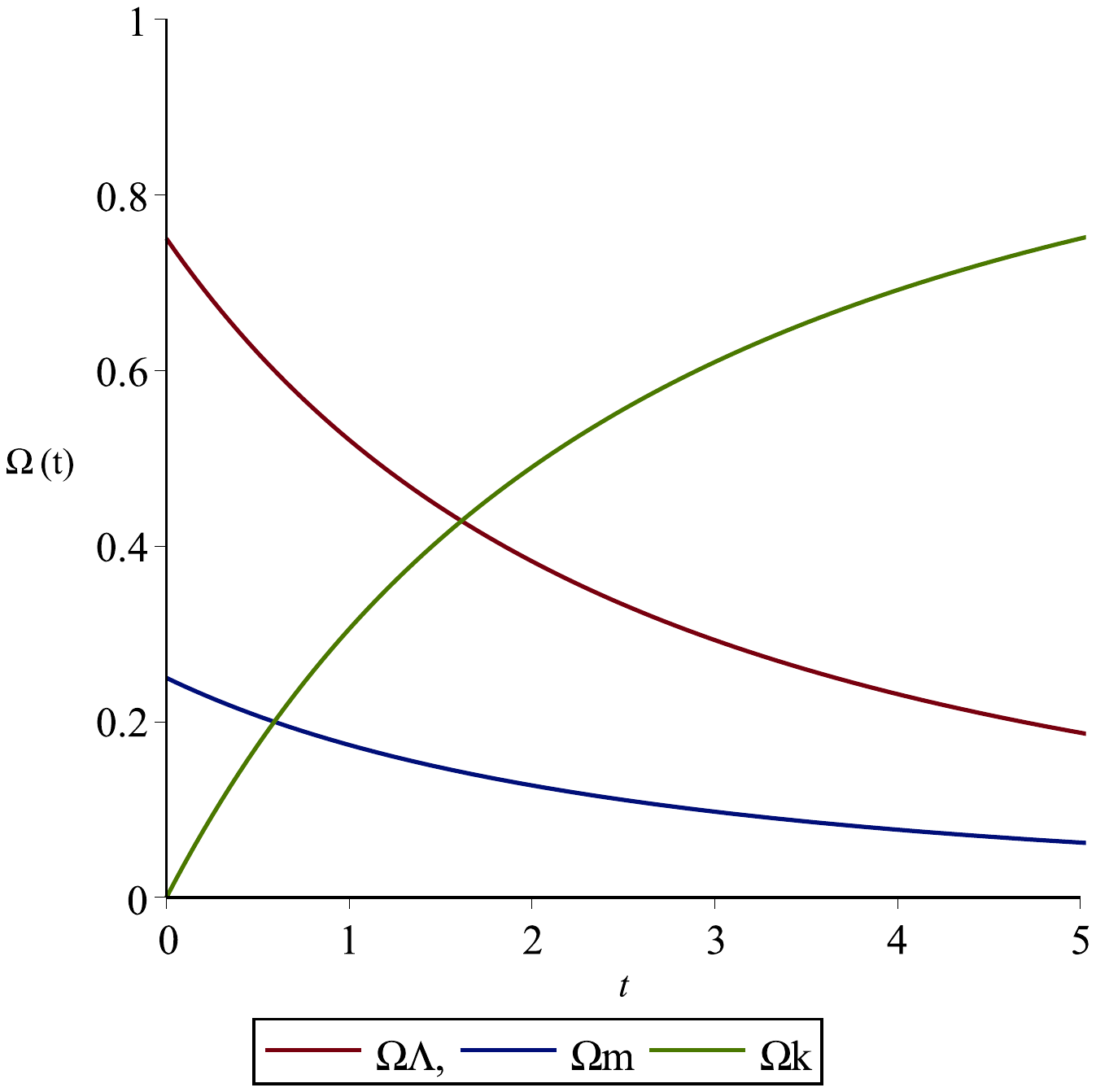}
		\vspace{-6cm}
		\caption{plot of $\Omega(t)$ versus $t$ for $k=-1$.}
	\end{center}
\end{figure} 

\begin{figure}
	\begin{center}
		\includegraphics[width=1.20\linewidth]{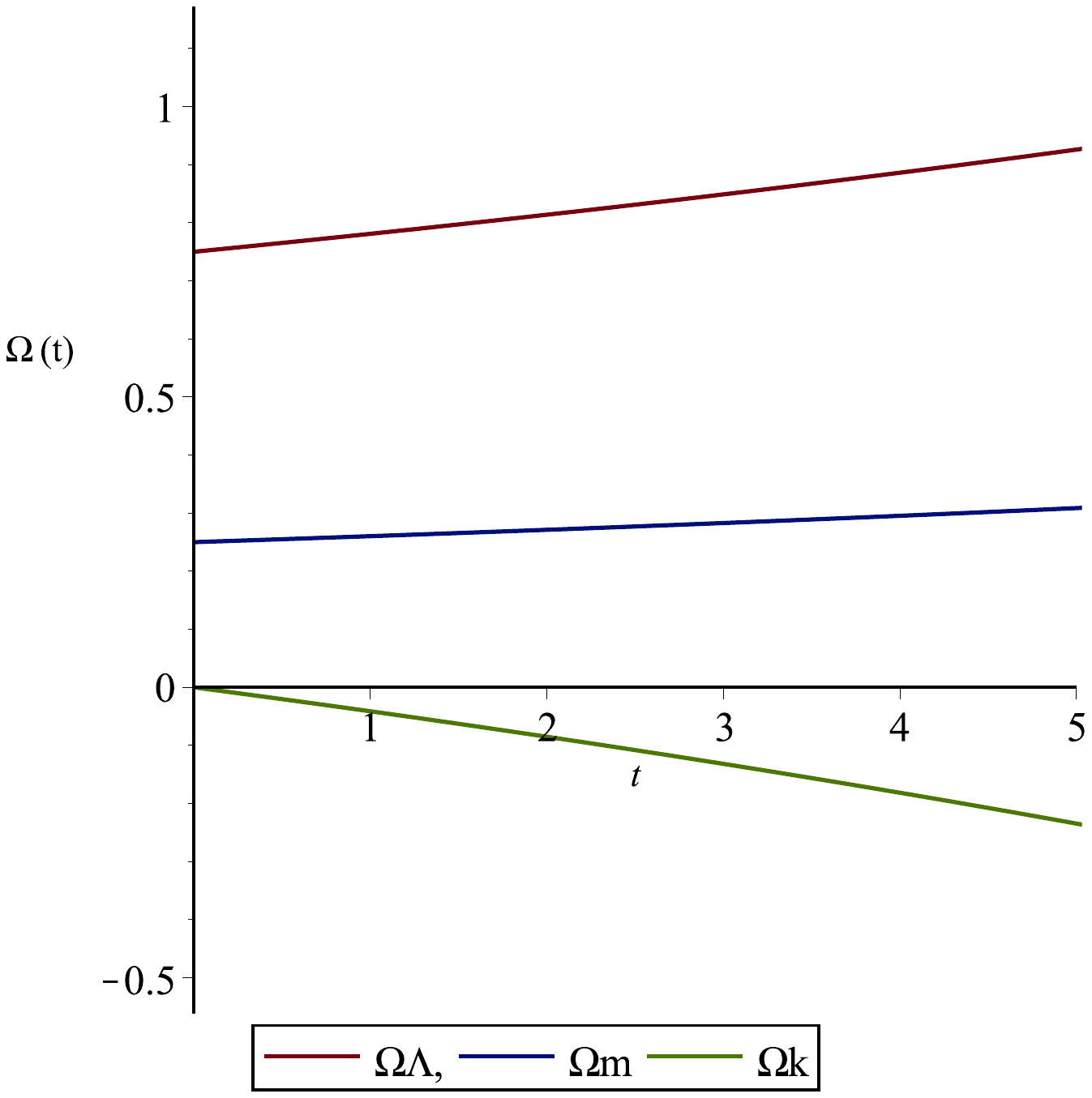}
		\vspace{-6cm}
		\caption{plot of $\Omega(t)$ versus $t$ for $k=+1$.}
	\end{center}
\end{figure}

\section{Constraints on the different parameters with latest observational results}
Although we deal with simple phenomenological models which are not dependent on any quantum field theory, different cosmological pictures can be reflected successfully.

Considering the cosmology of base-$\Lambda$-CDM, late-Universe parameters can be observed in ranges: Hubble constant $H_0=(67.4 \pm 0.5)$ km/s/Mpc; matter density parameter $\Omega_{m0}=0.315 \pm 0.007$\cite{planck}. Using the above ranges of $\Omega_{m0}$ the model parameters can be constraint as, $ -2.076\leq \beta_0 \leq -2.034$ and  $ 2.105\leq \gamma_0 \leq 2.247$.

 Present value of the cosmological constant $\Lambda_0$ can be obtained using the relation $\Lambda_0=3H_0^2\gamma_0\Omega_{m0}$. It lies within the range $ 0.9 \times 10^{-35} s^{-2} \leq \Lambda_0 \leq 1.042 \times 10^{-35} s^{-2}$, which is in sync with the observation \cite{planck}. We know the quintessence equation of state as $p_Q=\omega_Q \rho_Q$, $\omega_Q = -\Omega_{\Lambda}=-\gamma\Omega_{m}$. Using the above ranges we have, -0.724 $\leq \omega_Q \leq$ -0.648. This range is in good agreement with the accepted range of $\omega_Q$ which is -1 $\leq \omega_Q \leq$ -0.6\cite{balbi,snla,planck}, although, in either of our models we do not consider quintessence and present the range only as a qualitative check.

\section{Conclusion}
To summarize, the basic philosophy behind the present paper is to
generalize two phenomenological models. Explicit expressions of $a(t)$, $H(t)$, $\rho(t)$, $\Lambda(t)$ and also the parameter $\Omega(t)$ corresponding to matter, curvature and DE have been derived. Cosmic evolution of the Universe from the very early time to the late time has been discussed. 

The conclusions of the present work can be jot down as follows:

(i) The models $\Lambda\sim \frac{\ddot{a}}{a}$ and $\Lambda\sim \rho$ are equivalent for $k=\pm1$.

(ii) Both the models exhibit usual cosmological behaviour for early and late time Universe. Initially chosen values of the model parameters are found to be in good agreement with the observational data. 

(iii) Constraints on the different cosmological variables have been evaluated using our models and the results are in good agreement with the observational results.

\end{document}